\documentstyle[prl,aps,epsf]{revtex}
\newcounter{sub}
\newcounter{subeqn}[sub]
\begin{document}
\title
{ The Possible Role of R-modes in Post-glitch Relaxation of Crab }
\author{
Vahid Rezania $^{\dagger\ddagger}$and M. Jahan-Miri$\ddagger$ }

\address{~}
\address{
$\dagger$ Relativity and Cosmology Group, Division of Mathematics
and Statistics, \\ Portsmouth University, Portsmouth~PO1~2EG,
England }
\address{
$\ddagger$  Institute for Advanced Studies in Basic Sciences,
Zanjan 45195, Iran }
\address{vahid.rezania@port.ac.uk and
jahan@iasbs.ac.ir }

\date{\today}
\maketitle

\begin{abstract}
The loss of angular momentum through gravitational radiation,
driven by the excitation of r-modes, is considered in neutron
stars having rotation frequencies smaller than the associated
critical frequency. We find that for reasonable values of the
initial amplitudes of such pulsation modes of the star, being
excited at the event of a glitch in a pulsar, the total
post-glitch losses correspond to a negligible fraction of the
initial rise of the spin frequency in the case of Vela and the
older pulsars. However, for the Crab pulsar the same effect would
result, within a few months, in a decrease in its spin frequency
by an amount larger than its glitch-induced frequency increase.
This could provide an explanation for the peculiar behavior
observed in the post-glitch relaxations of the Crab.
\end{abstract}

~~~~~~~~~~~~~~{\bf keywords}  pulsars: individual (Crab Pulsar) --
stars: neutron -- stars: oscillations

\section{Introduction}
Many studies of the (nonradial) oscillation of neutron stars
concern the question of stability of the various modes realised in
these stars. The so-called CFS nonaxisymmetric instability, driven
by the emission of gravitational radiation, accompanies a very
rapid spinning down of a neutron star, at spin frequencies larger
than a certain critical frequency. Determination of such critical
frequencies for various modes and its observational implications
has been addressed by many authors (eg. Lindblom 1995). Also, for
the stable phase when the star rotates slower than the associated
critical frequency, possible observational manifestations of the
various modes (including those which arise from MHD,
superconductor, and superfluid effects) have been investigated
through a determination of the pulsation frequencies and the
damping times due to the different dissipation mechanisms (eg. Van
Horn 1980; McDermott, Van Horn \& Hansen 1988; Mendell 1991; Ipser
\& Lindblom 1991). A recently discovered curious feature of the
so-called r-modes is their generically unstable nature (Andersson
1998); i.e. the corresponding frequency for the onset of
CFS-instability equals zero, for some of these modes.
Nevertheless, a nonzero critical frequency $\Omega_{\rm c}$ is
still defined, in this case too, based on a comparison between the
``growth'' time of the mode (driven by gravitational radiation)
and its ``damping" time due to the viscosity. For \( \Omega >
\Omega_{\rm c} \), where $\Omega$ is the angular frequency of the
star, the unstable mode of oscillation grows in amplitude and a
significant decrease in $\Omega$ is predicted, also due to the
proportionality of the amplitude of gravitational waves with the
amplitude of the pulsation. However, when \(\Omega < \Omega_{\rm
c} \) the viscosity-driven damping is more efficient than the
growth, which is still driven by the emission of the gravity
waves. Much attention has again been paid to a determination of
the relevant $\Omega_{\rm c}$ and the observational implications
of the unstable phase in the case of r-modes (Andersson, Kokkotas
\& Stergioulas 1998; Madsen 1998; Levin 1998).

We are, however, concerned here with the consequences of r-mode
oscillation in neutron stars during their ``stable'' phase when
\(\Omega < \Omega_{\rm c} \). This is indeed expected to be the
relevant phase for all presently known pulsars (Andersson,
Kokkotas \& Schutz 1998). In particular, we will assume that
r-modes, originally absent in a pulsar, are excited
instantaneously at the event of a glitch which is observable as a
sudden spin-up of the star. The excited r-modes, which will be
subsequently damped effectively by the viscous effects, would
nevertheless result in certain amount of the angular momentum to
be taken away from the star by the gravitational waves. The total
loss of angular momentum until the modes are damped out is
computed, for assumed initial amplitudes of the excited r-modes.
The resulting decrease in the rotation frequency of the star is
compared with the (unrelaxed part of the) increase caused by the
glitch. The predicted effect is seen to be negligible in all
pulsars, except in the case of Crab which has also shown a
corresponding unique observational behavior, as will be discussed.
Normal modes of oscillation, including the r-modes, in a neutron
star might be excited at a glitch, which is observed as a sudden
rise of the spin frequency in radio pulsars (see, eg., Lyne 1995).
The excitation could be a result of the ``shaking'' of the star
due to a sudden transfer of angular momentum or a sudden variation
in the strength of the rotational coupling, between its various
components, or due to a sudden change of its moment of inertia
accompanying structural changes. All these effects are commonly
considered as possibilities which might occur at a glitch (eg.
Blandford 1992). It is further noted that the special role
attributed here to the r-modes during a post-glitch relaxation has
not to do with their distinctive unstable property, as was
indicated earlier. The other (say, p-, f-, g-) modes of
oscillation in a neutron star could have, in principle, played the
same role as well. The distinction lies, however, in the
associated damping times. The damping time for the r-modes (in the
case of Crab pulsar) is comparable to that of the observed
post-glitch relaxation. In contrast, the damping time for the
other modes is either too short or too long by many orders of
magnitudes (eg., McDermott et~al 1984, Kokkotas and Schmidt 1999).
Hence the total loss of angular momentum due to gravitational
radiation by the excited modes, during intervals between
successive glitches in, say, Crab Pulsar turns out to be
negligible, for all the other modes except for the r-modes (see
Fig.~1 below).

\section{The Predicted Effect}
In order to estimate the effect of the r-mode instability, in a
``stable'' neutron star, on its post-glitch relaxation, we have
used the model described by Owen et.~al. (1998). The total angular
momentum of a star is parameterized in terms of the two degrees of
freedom of the system. One, is the uniformly rotating equilibrium
state which is represented by its angular velocity $\Omega_{\rm
eq}$. The other, is the excited r-mode that is parameterized by
its magnitude $\alpha$ which is bound to an upper limiting value
of $\alpha=1$, in the linear approximation regime treated in the
model. Thus, the total angular momentum $J$ of the star is written
as a function of the two parameters $\Omega_{\rm eq}$ and
$\alpha$: \stepcounter{sub}
\begin{equation}
J=I_{\rm eq} \Omega_{\rm eq} + J_c,
\end{equation}
where $I_{\rm eq}=\tilde{I} M R^2$ is the moment of inertia of the
equilibrium state, and $J_c=-{3 \over 2} \tilde{J} \alpha^2
\Omega_{\rm eq} M R^2$ is the canonical angular momentum of the
$l=2$ r-mode, which is negative in the rotating frame of the
equilibrium star. The dimensionless constants $\tilde{I}$ and
$\tilde{J}$ depend on the detailed structure of the star, and for
the adopted $n=1$ polytropic model considered have values
$\tilde{I}=0.261$ and $\tilde{J}=0.01635$. Also $R=12.54$~km and
$M=1.4 \ M_\odot$ are the assumed radius and mass of the star, for
the same polytropic model.

Eq.~1 above implies that an assumed instantaneous excitation of
r-modes at a glitch would cause a sudden increase in $\Omega_{\rm
eq}$. For definiteness, we define the ``real'' observable rotation
frequency $\Omega$ of the star as $\Omega = {J \over I}$, where
$I$ is the moment of inertia of the real star. The two are equal,
$\Omega =\Omega_{\rm eq}$, in the absence of the r-modes, ie.
before the excitation of the modes at a glitch and after the modes
are damped out. If there were no loss of angular momentum (by
gravitational radiation) accompanying the post-glitch damping (by
viscosity) of the modes $\Omega_{\rm eq}$ would recover its
extrapolated pre-glitch value; ie. its initial rise would be
compensated exactly. However due to the net loss of angular
momentum by the star, the post-glitch decrease of $\Omega_{\rm
eq}$ {\it overshoots} its initial rise. The negative offset
between the values of $\Omega_{\rm eq}$ before the excitation of
the modes and after they are damped out is the quantity of
interest for our discussion. The question of whether the
instantaneous rise in the value of $\Omega_{\rm eq}$ at a glitch,
due to the excitation of the r-modes, is  observable or not is a
separate problem, and its resolution would have no consequence for
the net loss of angular momentum from the star which is the
relevant quantity here. It is noted that the distinction between
$\Omega$ and $\Omega_{\rm eq}$, in the presence of modes, is
quantitatively negligible, in all cases of interest, and is
usually disregarded. Also, one might dismiss an {\em increase} in
$\Omega_{\rm eq}$ as implied by Eq.~1 to be observable as a
spin-up of the star since for an inertial outside observer the
r-modes rotate in the prograde direction and their excitation
should result, if at all, in a {\it spin-down} of the star.
Moreover, an excitation of the r-modes should not result, by
itself, in any real change of the rotation frequency of the star
at all. Because one could not distinguish two physically separate
parts of the stellar material such that the two components of
angular momentum in Eq.~1 may be assigned to the two parts
separately.

The time evolution of the quantities $\alpha$ and $\Omega_{\rm
eq}$ may be determineded from the coupled equations (Owen et.~al.
1998): \stepcounter{sub} \stepcounter{subeqn}
\begin{eqnarray}
&&\frac{{\rm d}\Omega_{\rm eq}}{{\rm d}t}= -\frac{2\Omega_{\rm
eq}}{\tau_{\rm v}}\frac{\alpha^2 Q}{1+\alpha^2 Q},\\
\stepcounter{subeqn} &&\frac{{\rm d}\alpha}{{\rm d}t}=
-\frac{\alpha}{\tau_{\rm gr}}-\frac{\alpha}{\tau_{\rm v}}
\frac{1-\alpha^2 Q}{1+\alpha^2 Q} ,
\end{eqnarray}
where $Q= {3 \over 2}{\tilde{J} \over \tilde{I}} = 0.094$, for the
adopted equilibrium model of the star. The two timescales
$\tau_{\rm v} (>0)$ and $\tau_{\rm gr} (<0)$ are the viscous
damping and gravitational growth timescales, respectively. The
viscous time has two contributions from the shear and bulk
viscosities with corresponding timescales $\tau_{\rm sv}$ and
 $\tau_{\rm bv}$, respectively. The overall ``damping''
timescale $\tau$ for the mode, which is a measure of the period
over which the excited mode will persist, is defined as
\stepcounter{sub}
\begin{equation}
\frac{1}{\tau}=\frac{1}{\tau_{\rm v}}+\frac{1}{\tau_{\rm gr}}=
\frac{1}{\tau_{\rm sv}}+\frac{1}{\tau_{\rm bv}}+\frac{1}{\tau_{\rm
gr}}
\end{equation}
Following Owen et.~al. (1998) we use \(\tau_{\rm sv}=2.52\times
10^8 ({\rm s}) T_9^2 \), \(\tau_{\rm bv}=4.92\times 10^{10} ({\rm
s}) T_9^{-6} \Omega_3^{-2} \), and \(\tau_{\rm gr}=-1.15\times
10^6 ({\rm s}) \Omega_3^{-6}\), where $T_9$ is the temperature,
$T$, in units of $10^9$~K, and $\Omega_3$ is in units of $10^3
{\rm rad \ s}^{-1}$. The above expression for $\tau$ which we have
used for our following calculations does not however include the
role of superfluid mutual friction in damping out the
oscillations. We have further taken into account that effect by
including also a damping time $\tau_{\rm mf} =4.28\times 10^8
\Omega_3^{-5}$ sec, due to the mutual friction (Lindblom and
Mendell 1999) in the calculation of $\tau$. The effect of the
mutual friction is nevertheless seen to be negligible and the
computed curves shown below remain almost the same even when the
effect due to the mutual friction is included.

By integrating Eqs~2, numerically, for a given initial value of
$\alpha$, one may therefore follow the time evolution of $\alpha$
and $\Omega_{\rm eq}$ which together with Eq.~1 determine the time
evolution of the total angular momentum, $J$, and hence the time
evolution of $\Omega$. Fig.~1 shows the computed time evolution
for the absolute value of the resulting (negative) fractional
change $\Delta \Omega \over \Omega$ in the spin frequency
(Fig.~1a) and also the change $\Delta \dot \Omega \over\dot
\Omega$ in the spin-down rate of the star (Fig.~1b), starting at
the glitch epoch which corresponds to time $t=0$. The results in
Fig.~1 are for a choice of an initial value of \(\alpha_0=0.04 \),
and for the assumed values of $T$ and $\Omega$ corresponding to
the Crab and Vela pulsars, as indicated. Fig.~1a shows that for
the same amplitude of the r-modes assumed to be excited at a
glitch the resulting loss of angular momentum through
gravitational radiation would be much larger in Crab than in Vela,
ie. by more than 3 orders of magnitudes. (Note that the curve for
Vela in Fig.~1a represents the results after being multiplied by a
factor of $10^3$.) Furthermore, for the adopted choice of
parameter values, the magnitude of the corresponding decrease in
$\Omega$ for the Crab, is \(|{\frac{\Delta\Omega}{\Omega}}| \sim
10^{-7} \) (Fig.~1a). The observational consequence of such an
effect would nevertheless be closely similar to what has been
already observed during the post-glitch relaxations of, {\it
only}, the Crab pulsar.

Before proceeding further with Crab, we note that the post-glitch
effects of excitation of r-modes would however have not much
observational consequences for the Vela, and even more so for the
older pulsars, which are colder and rotate more slowly. This has
two, not unrelated, reasons: in the older pulsars r-modes a) are
damped out faster (ie. have smaller values of $\tau$), and b)
result in less gravitational radiation. The dependence of $\tau$
on the stellar interior temperature is shown in Fig.~2a. For the
colder, i.e. older, neutron stars the r-modes are expected to die
out very fast. The damping timescale for a pulsar with a period \(
P \sim 1 {\rm s} \), being colder than $10^8$~K, could be as short
as a few hours (Fig.~2a), and r-modes would have been died out at
times longer than that after a glitch. For the hot Crab pulsar, on
the other hand, r-modes are expected to persist for 2-3 years
after they are excited, say, at a glitch. The value of $\tau$
decreases for older pulsars due to both their longer periods as
well as lower temperatures, but the effect due to the latter
dominates by many orders of magnitudes, for the standard cooling
curves of neutron stars (Urpin et. al. 1993). The second reason,
ie. the loss of angular momentum being negligible in older
pulsars, was already demonstrated in Fig.~1a, by a comparison
between Crab and Vela pulsars. We have verified it also for the
case of pulsars older than Vela. It may be also demonstrated
analytically from Eqs~2, in the limit of $\alpha^2 Q << 1$. The
initial increase in $\Omega_{\rm eq}$ due to excitation of r-modes
with a given initial amplitude $\alpha_0$ is seen from Eq.~1 to be
$|{\Delta\Omega_{\rm eq}\over\Omega_{\rm eq}}|_0= \alpha_0^2 Q$.
The subsequent damping of the modes result in secular decrease in
$\Omega_{\rm eq}$, and the total decrease at large
$t\rightarrow\infty$ would be $|{\Delta\Omega_{\rm
eq}\over\Omega_{\rm eq}}|_\infty \sim {\tau \over \tau_{\rm
v}}\alpha_0^2 Q$, which is true for $|{\Delta\Omega_{\rm eq}| <<
\Omega_{\rm eq}}$. Note that in the absence of gravitational
radiation losses (ie. ${1 \over \tau_{\rm gr}}=0; \tau = \tau_{\rm
v}$) the total decrease would be the same as the initial increase,
which is expected for the role of viscous damping alone. The
difference between these two changes (total decrease minus initial
increase) in $\Omega_{\rm eq}$ would correspond to the total loss
of angular momentum from the star, hence to the net decrease in
its observable rotation frequency, ie. \stepcounter{sub}
\begin{equation}
|{\Delta\Omega \over\Omega}|_\infty= {\tau_{\rm v} \over
|\tau_{\rm gr}|} \alpha_0^2 Q
\end{equation}
which is valid in the limit of ${\tau_{\rm v} \over |\tau_{\rm
gr}|} << 1$. Fig.~2b shows the dependence of the quantity
${\tau_{\rm v} \over |\tau_{\rm gr}|}$ on the stellar rotation
frequency, and also on its internal temperature. While for the
Crab ${\tau_{\rm v} \over |\tau_{\rm gr}|} \sim 10^{-3}$, however
its value is much less for the older pulsars, due to both their
lower $\Omega$ as well as lower $T$ values. The dependence on the
temperature is however seen to be much less than that on the
rotation frequency, in contrast to the dominant role of the
temperature in determining the value of the total damping time
$\tau$, as indicated above. As is seen in Fig.~2b, for the Vela
${\tau_{\rm v} \over |\tau_{\rm gr}|} < 10^{-7}$, which means a
maximum predicted value of \(|{\Delta\Omega \over\Omega}|_\infty
<10^{-8} \), even for the large values of $\alpha_0 \sim 1$. This
has to be contrasted with the glitch induced values of
\(|{\Delta\Omega \over\Omega}| \sim1 0^{-6} \) in Vela, which
shows the insignificance of the role of r-modes in its post-glitch
behaviour.

\section{The Crab}
The observed post-glitch relaxation of the Crab pulsar has been
unique in that the rotation frequency of the pulsar is seen to
decrease to values $less$ than its pre-glitch extrapolated values
(Lyne et. al. 1993). So far, two mechanisms have been suggested to
account for the observed excess loss of angular momentum during
post-glitch relaxations of the Crab. The first mechanism, in the
context of the vortex creep theory of Alpar et.~al. (1984, 1996),
invokes generation, at a glitch, of a so called ``capacitor''
region within the pinned superfluid in the crust of a neutron
star, resulting in a permanent decoupling of that part of the
superfluid. Nevertheless, this suggestion has been disqualified
(Lyne et. al. 1993) since the moment of inertia required to have
been decoupled permanently in such regions during the past history
of the pulsar is found to be much more than that permitted for
{\it all} of the superfluid component in the crust of a neutron
star. In another attempt, Link et.~al. (1992) have attributed the
excess loss of angular momentum to an increase in the
electromagnetic braking torque of the star, as a consequence of a
sudden increase, at the glitch, in the angle between its magnetic
and rotation axes. As they point out, such an explanation is left
to future observational verification since it should also
accompany other observable changes in the pulsar emission, which
have not been detected, so far, in any of the resolved glitches in
various pulsars. Moreover, the suggestion may be questioned also
on the account of its long-term consequences for pulsars, in
general. Namely, the inclination angle would be expected to show a
correlation with the pulsar age, being larger in the older pulsars
which have undergone more glitches. No such correlation has been
deduced from the existing observational data.   Also, and even
more seriously, the assumption that the braking torque depends on
the inclination angle is in sharp contradiction with the common
understanding of pulsars spin-down process. The currently inferred
magnetic field strengths of all radio pulsars are in fact based on
the opposite assumption, namely that the torque is independent of
the inclination angle. The well-known theoretical justification
for this, following Goldreich \& Julian (1969), is that the torque
is caused by the combined effects of the magnetic dipole radiation
and the emission of relativistic particles, which compensate each
other for the various angles of inclination (see, eg., Manchester
\& Taylor 1977; Srinivaran 1989).

The excitation of r-modes at a glitch and the resulting emission
of gravitational waves could, however, account for the required
``sink'' of angular momentum in order to explain the peculiar
post-glitch relaxation behavior of the Crab pulsar. As is shown in
Fig.~1, for values of $ \alpha_0 \geq 0.04 $ the predicted time
evolution of $\Delta \Omega \over \Omega$ and $\Delta\dot \Omega
\over\dot\Omega $ during the 3--5 years of the inter-glitch
intervals in Crab, might explain the observations. That is, the
predicted total change in the rotation frequency of the star,
$|{\Delta\Omega\over\Omega}|$,  is much larger than the
corresponding jump $\frac{\Delta\Omega}{\Omega}\sim 10^{-8}$ at
the glitch, which explains why the post-glitch values of $\Omega$
should fall below that expected from an extrapolation of its
pre-glitch behavior. Also, the predicted values of ${\Delta \dot
\Omega \over\dot\Omega} \sim 10^{-4}$, after a year or so
(Fig.~1b), are in good agreement with the observed persistent
shift in the spin-down rate of the Crab (Lyne et.~al. 1995).
The predicted increase in the spin-down rate would be however
diminished as the excited modes at a glitch are damped out,
leaving a permanent negative offset in the spin frequency. Hence
the above so-called persistent shift in the spin-down rate of the
Crab may be explained in terms of the effect of r-modes, as long
as it persists during the inter-glitch intervals of 2-3 years. It
may be noted that a permanent persistent shift in the spin-down rate
at a glitch may be caused by a sudden decrease in the moment of
inertia of the star. However this effect could not, by itself,
result in the observed {\it negative} offset in the spin
frequency, at the same time.

The same mechanism would be expected to be operative during the
post-glitch relaxation in the other {\it colder} and {\it slower}
pulsars, as well. However, for the similar values of $\alpha_0$,
ie. the same initial amplitude of the excited modes, the effect is
not expected to become ``visible'' in the older pulsars.
Particularly, for the Vela its initial jump in frequency at a
glitch, $\frac{\Delta\Omega}{\Omega}\sim 10^{-6}$, is seen from
Fig.~1a to be much larger (ie. by some four orders of magnitudes)
than that of the above effect due to the r-modes. In other words,
while the predicted loss in the stellar angular momentum due to
the excitation of r-modes result in a negative $\Delta \Omega
\over \Omega$ which, in the case of Crab, overshoots the initial
positive jump at a glitch, however for the Vela and older pulsars
it comprises only a negligible fraction of the positive
glitch-induced jump. A more detailed study should, however, take
into account the added complications due to internal relaxation of
various components of the star, which is highly model dependent.
The observed initial rise in $\Omega$ need not be totally
compensated for by the losses due to r-modes which we have
discussed, since part of it could be relaxed internally (by a
transfer of angular momentum between the ``crust'' and other
components, and/or temporary changes in the effective moment of
inertia of the star) even in the absence of any real sink for the
angular momentum of the star. Such considerations would not only
leave the above conclusions valid but also allow for even smaller
values of the initial amplitude of the excited modes, compared to
our presently adopted value of $\alpha_0 \sim 0.04$. The suggested
effect of the r-modes in the post-glitch relaxation of pulsars
should be understood as one operating in addition to that of the
internal relaxation which is commonly invoked. While the latter
could account {\it only} for a relaxation back to (or higher than)
the extrapolated pre-glitch values of the spin frequency, the
additional new effect due to the r-modes may explain the {\it
excess} spin-down to the lower values, as is observed in the Crab
pulsar.

It is further noted that the above estimates are for an adopted
value of \( Q = 9.4\times 10^{-2}\), which corresponds to the
particular choice of the polytropic model star. Differences in the
structure among pulsars, in particular between Crab and Vela,
which have also been invoked in the past (see, eg., Takatsuka \&
Tamagaki 1989), could be further invoked to find a better
agreement with the data for the above effect due to r-modes as
well. Also, the initial amplitude of the excited modes need not be
the same in all pulsars. It is reasonable to assume that in a
hotter and faster rotating neutron star, as for the Crab, larger
initial amplitudes, ie. larger values of $\alpha_0$, are realised
than in the colder--slower ones.

The quantity $\alpha_0$ is however a free parameter in our
calculations for which we have chosen a value  $\alpha_0 \sim
0.04$ that result in an appreciable loss of angular momentum for
the Crab pulsar. Nevertheless, for the {\it same} choice of the
value of $\alpha_0$ the effect of r-modes would be different
between the Crab and the other pulsars, in agreement with the
observations. The associated energy of the excited r-modes, in the
rotating frame, is \( \tilde{E} = {1\over 2} \alpha_0^2 \tilde{J}
\Omega^2 M R^2\), which result in \( \frac{\tilde{E}}{E_{\rm rot}}
\sim 10^{-4} \) for a value of $\alpha_0 \sim 0.04$, where $E_{\rm
rot} = \frac{I \Omega^2}{2}$ is the rotational energy of the star.
In contrast, for the energy \(\Delta E \sim I \Omega \Delta
\Omega\) released at a glitch, associated with a glitch of a
change $\Delta \Omega$ in the rotation frequency, \( \frac{\Delta
E}{E_{\rm rot}} \sim \frac{\Delta \Omega}{\Omega}\) which is much
smaller than the above even for the giant glitches of the Vela
pulsar. We note however that the excitation of the oscillation
modes is a separate process taking place in the liquid core of the
star, and need not be energetically comparable to the energy
transfer/dissipation involved in a glitch. Furthermore, as
indicated earlier, one should note that the distinction between
modes and the equilibrium star, as in Eq.~1, is only a
mathematical convenience and no real transfer of angular momentum
takes place from one component into another. Hence the associated
energy of the modes need not have any definite relation to that of
the glitch process which is accompanied by a real transfer of
angular momentum between different components of the star.
Unfortunately, the existing theory does not provide any
prescription for determining the initial amplitude of the excited
modes, for any assumed cause of it, say, at a glitch.

We are thankful to the referee for the valuable comments and
criticisms which helped to improve the manuscript. A helpful
correspondence from K. D. Kokkotas is also gratefully
acknowledged.  VR acknowledges the support of a Royal Society
grant and thanks the Portsmouth Relativity and Cosmology Group for
hospitality.\\
\\

{\bf References}\\
\begin{itemize}
\item[] Alpar M. A., Anderson P. W., Pines D., Shaham J.,
        1984, ApJ, 276, 325
\item[] Alpar M. A., Chau H. F., Cheng K. S., Pines D.,
        1996, ApJ, 459, 706
\item[] Andersson N., 1998, ApJ, 502, 708
\item[] Andersson N., Kokkotas K. D., Schutz B. F.,
        1999, ApJ, 510, 846
\item[] Andersson N., Kokkotas K. D., Stergioulas N.,
        1999, ApJ, 516, 846
\item[] Blandford R. D., 1992, Nat., 359, 675
\item[] Goldreich P., Julian W. H., 1969, ApJ, 157, 869
\item[] Ipser J. R., Lindblom L., 1991, ApJ, 373, 213
\item[] Kokkotas D. K., Schmidt B. G., 1999, e-print,
gr-qc/9909058
\item[] Levin Y., 1999, ApJ, 517, 328
\item[] Lindblom L., 1995, ApJ, 438, 265
\item[] Lindblom L., Mendell G., 1999, e-print, gr-qc/9909084
\item[] Link B., Epstein R. I., Baym G., 1992, ApJ, 390, L21
\item[] Lyne A. G., 1995,, in Alpar M. A.,
         Kizilo\v{g}lu \"{U}., van Paradijs J. (eds), The
         Lives of the Neutron Stars. Kluwer, Dordrecht, p. 167
\item[] Lyne A. G., Pritchard R. S., Smith F. G., 1993,
        MNRAS, 265, 1003
\item[] Madsen J., 1998, Phys. Rev. Lett., 81, 3311
\item[] Manchester R. N., Taylor J. H., 1977, ``Pulsars'', Freeman,
              San Fransisco
\item[] McDermott P. N., Svaedoff M. P., Van Horn H. M., Zweibel E. G., Hansen C. J.,  1984, ApJ,  281, 746
\item[] McDermott P. N., Van Horn H. M., Hansen c. J., 1988,
        ApJ, 325, 725
\item[] McKenna J., Lyne A. G., 1990, Nat., 343, 349
\item[] Mendell G., 1991, ApJ, 380, 515
\item[] Owen B. J., Lindblom L., Cutler C., Schutz B. F., Vecchio A.,
Andersson N., 1998, Phys. Rev. D, 58, 084020
\item[] Srinivaran G., 1989, Astronomy \& Astrophysics Review, 1, 209
\item[] Takatsuka T. Tamagaki R., 1989, Progress in Theoretical Physics,
         Vol. 82, No. 5, 945
\item[] Urpin V. A., Van Riper K. A., 1993, ApJ, 411, L87
\item[] Van Horn H. M., 1980, ApJ, 236, 899
\end{itemize}

\clearpage

\null
\noindent
{\bf \large Figure Captions:} \\

\vspace{1cm}
\noindent

{\bf Fig. 1a-}  The post-glitch time evolution of the absolute
value of the fractional change in the spin frequency of a pulsar,
caused by its loss of angular momentum due to gravitational waves
driven by the r-modes that are assumed to be excited at the glitch
epoch, $t=0$, with an initial amplitude of $\alpha_0=0.04$. The
two curves correspond to assumed values of $T_9=0.3$ and $\Omega_3
=0.19$, for the Crab ({\it thick} line), and $T_9=0.2$ and
$\Omega_3 =0.07$, for the Vela ({\it thin} line). Note that the
curve for Vela represents the results {\bf after being multiplied}
by a factor of $10^3$.

\vspace{1cm}
\noindent
{\bf Fig. 1b-} Time evolution of the fractional change
in the spin-down rate of a pulsar, caused by its loss of
angular momentum due to the excitation of r-modes at $t=0$.
A value of $\dot\Omega= 2.4 \times 10^{-9}$rad s$^{-2}$, and
other parameter values same as in Fig.~1a for the Crab have
been assumed.

\vspace{1cm}
\noindent
{\bf Fig. 2a-} The dependence of the total damping
timescale of r-modes on the internal
temperature of a neutron star.
Parameter values same as for the Crab in Fig.~1a have
been assumed.

\vspace{1cm}
\noindent
{\bf Fig. 2b-} The dependence of the quantity
${\tau_{\rm v} \over |\tau_{\rm gr}|}$, which is a
measure of the net post-glitch decrease in the rotation
frequency, on the rotation frequency of a neutron star.
The two curves are to show the dependence on the temperature,
where $T_9=0.3$ ({\it bare} line) and
$T_9=0.2$ ({\it dotted} line) have been used.

\clearpage

\begin{figure}
\epsfxsize=4.4in \epsfysize=4.0in \epsffile{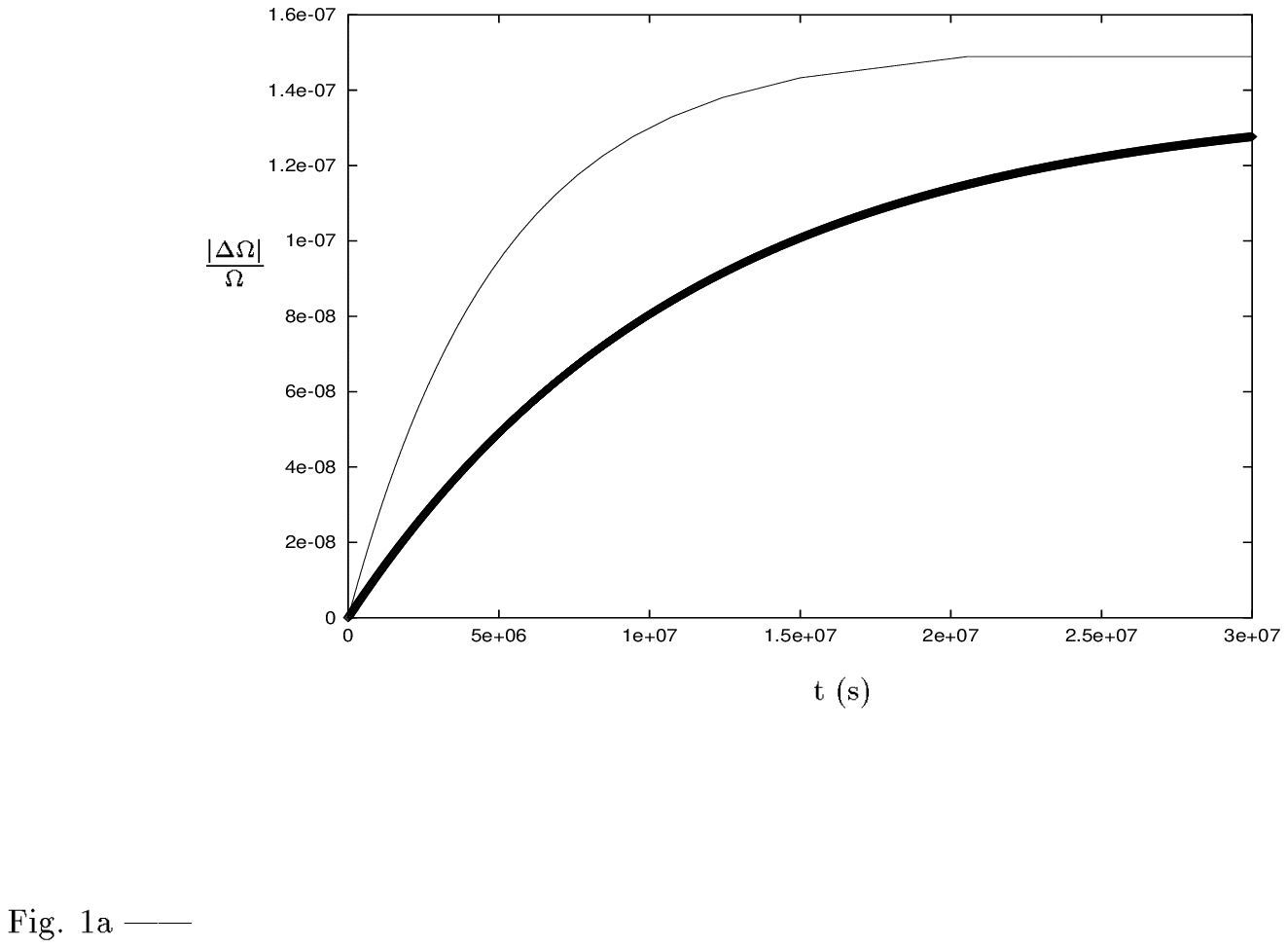}
\end{figure}

\begin{figure}
\epsfxsize=4.4in \epsfysize=4.0in \epsffile{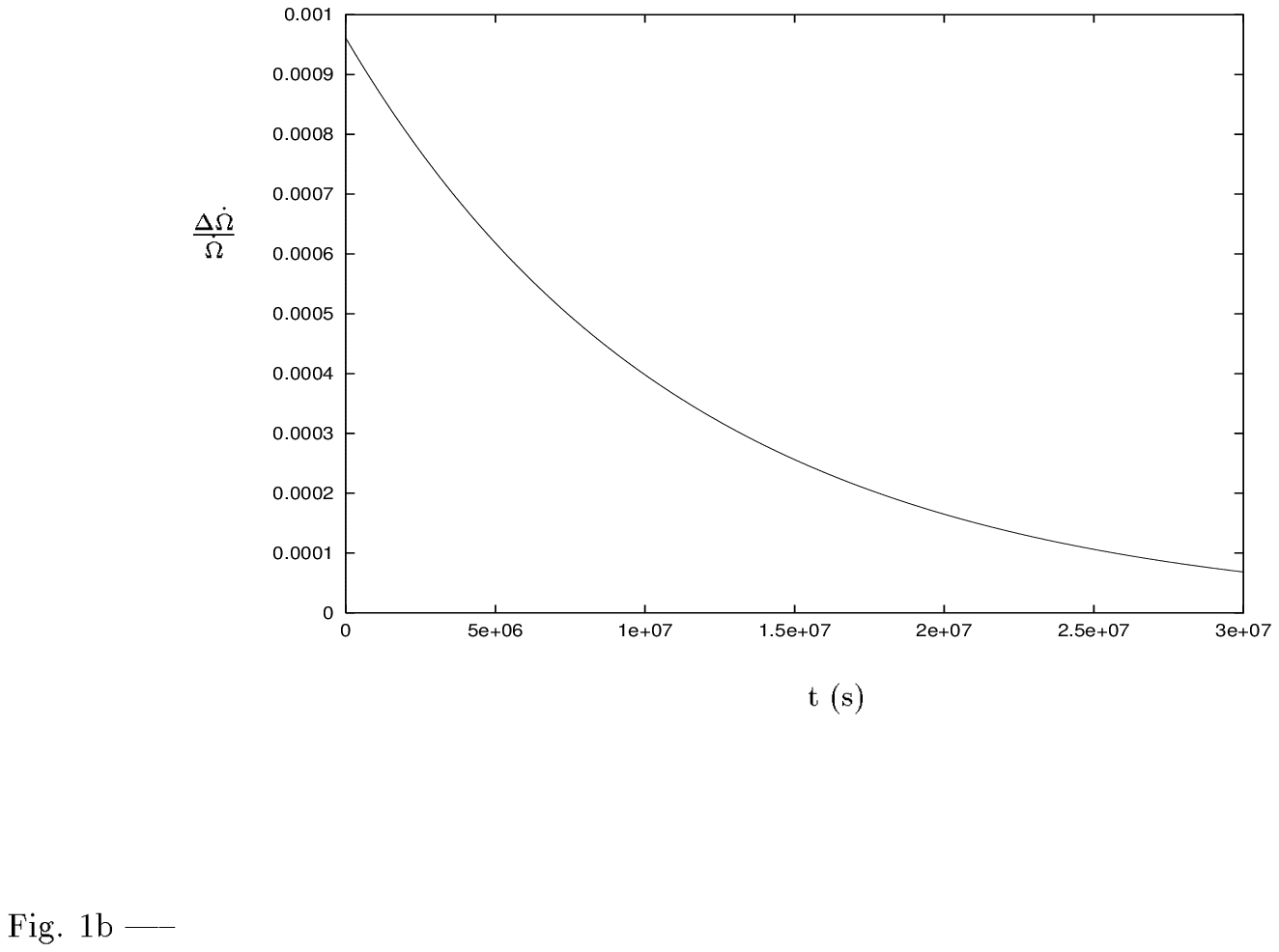}
\end{figure}

\begin{figure}
\epsfxsize=4.4in \epsfysize=4.0in \epsffile{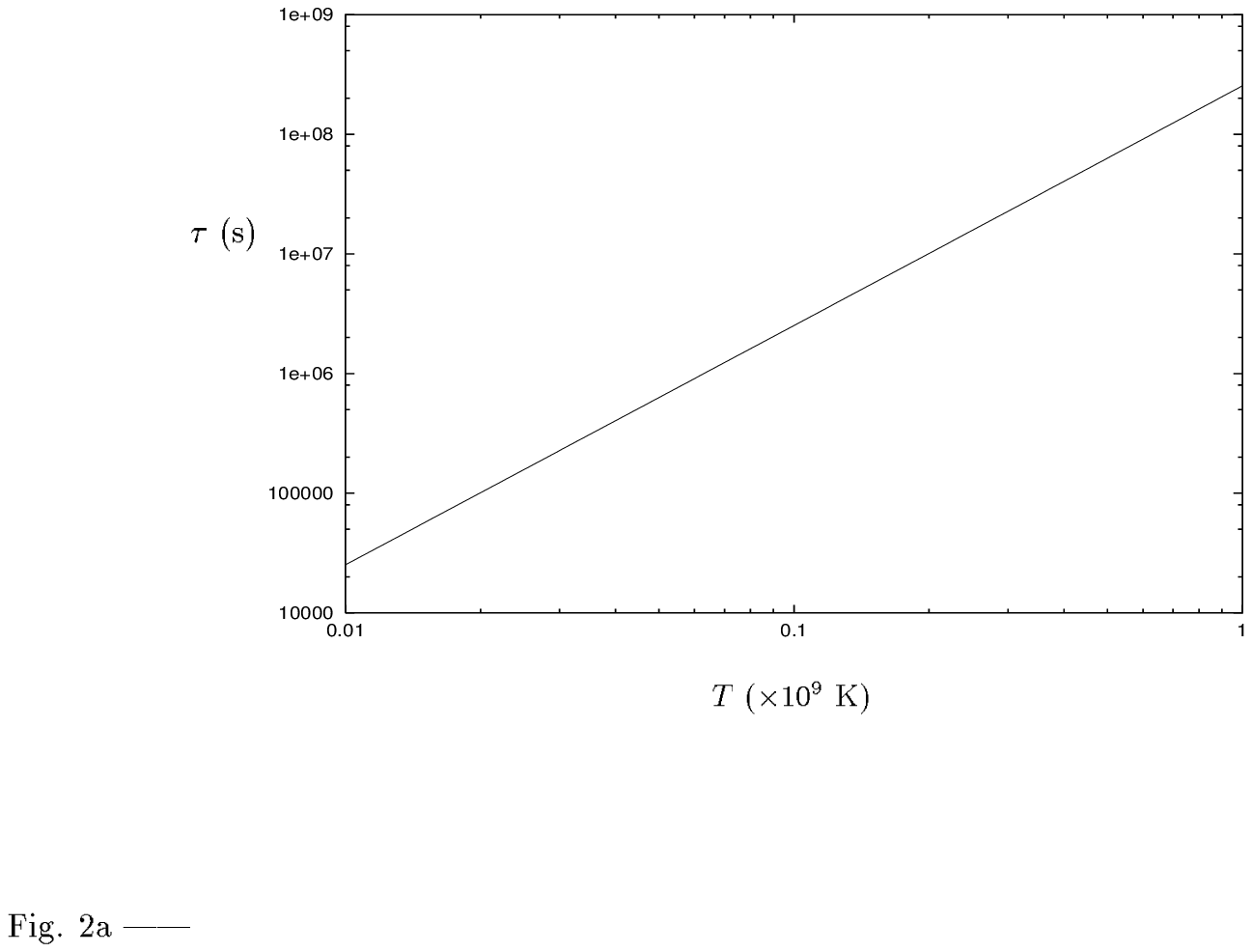}
\end{figure}

\begin{figure}
\epsfxsize=4.4in \epsfysize=4.0in \epsffile{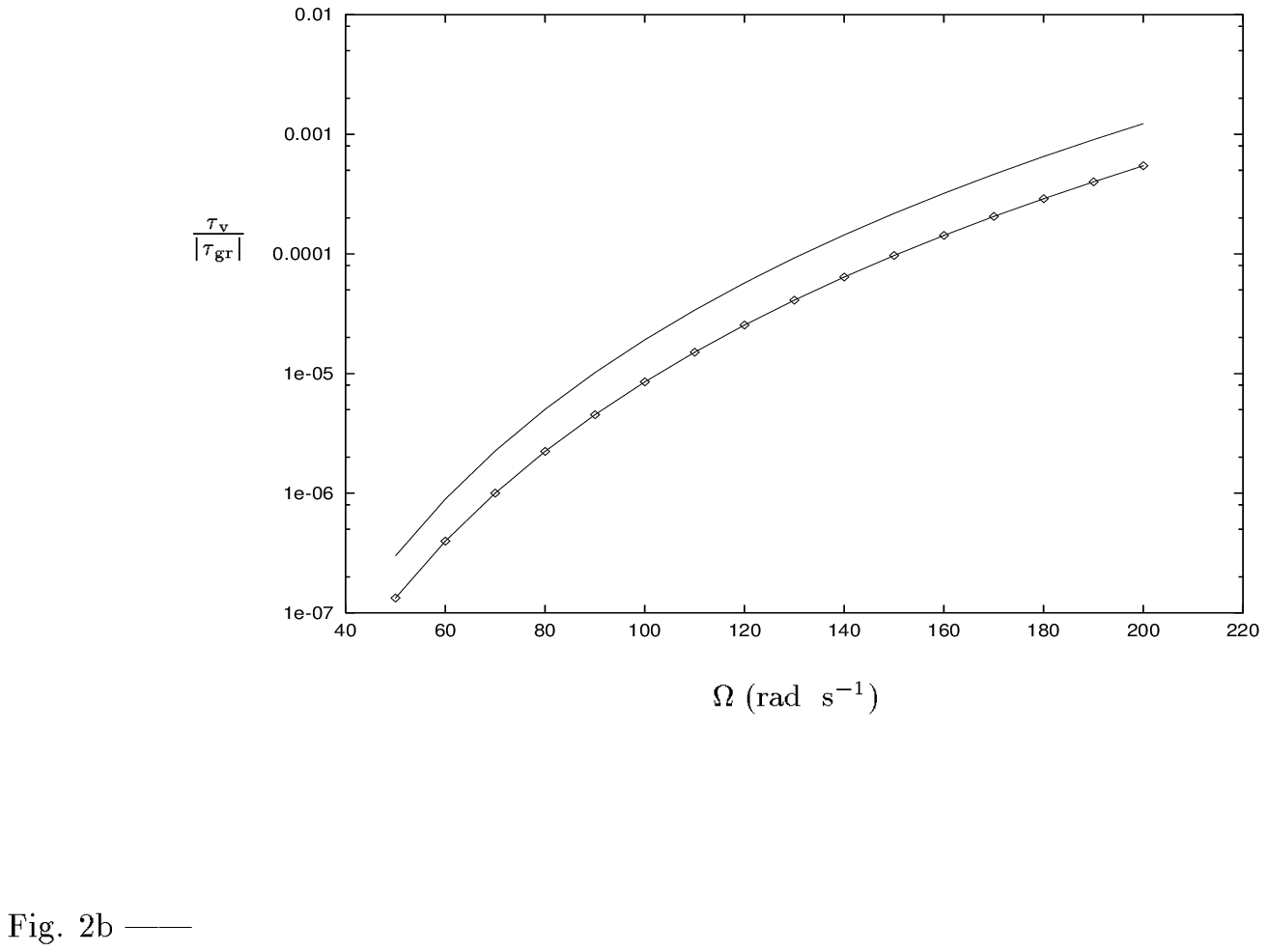}
\end{figure}
\end{document}